\documentclass[a4paper,12pt]{article}
\usepackage{epsfig, amsmath, fullpage, amssymb}

\begin{document}

\def\newtheorems{
                 \newtheorem{theorem}{Theorem}[section]
                 \newtheorem{cor}[theorem]{Corollary}
                 \newtheorem{prop}[theorem]{Proposition}
                 \newtheorem{property}[theorem]{Property}
                 \newtheorem{lemma}[theorem]{Lemma}
                 \newtheorem{defn}[theorem]{Definition}
                 \newtheorem{claim}[theorem]{Claim}
                 \newtheorem{sublemma}[theorem]{Sublemma}
                 \newtheorem{example}[theorem]{Example}
                 \newtheorem{remark}[theorem]{Remark}
                 \newtheorem{question}[theorem]{Question}
                 \newtheorem{conjecture}{Conjecture}[section]}
                 \newtheorem{problem}{Problem}[section]
                 \newtheorem{notation}{Notation}[section]
\newtheorems
\newcommand{\proof}{\par\noindent{\bf Proof:}\quad}
\newcommand{\solution}{\par\noindent{\bf Solution:}\quad}
\newcommand{\minla}{Minimum Linear Arrangement problem }

\title{The Minimum Linear Arrangement Problem on Proper Interval Graphs}
\author{Ilya Safro\footnote{This work is done as a part of M.Sc. thesis \cite{safro}}}

\maketitle


\abstract{We present a linear time algorithm for the minimum linear arrangement problem on proper interval graphs. The obtained ordering is a 4-approximation for general interval graphs.}
\section{Preliminaries}

\par Let $\mathcal{F}$ be a family of nonempty sets. The {\it intersection graph} of $\mathcal{F}$ is obtained by representing each set in $\mathcal{F}$ by a vertex and connecting two vertices by an edge if and only if their corresponding sets intersenct. The intersection graph of a family of intervals on a linearly ordered set (like the real line) is called an {\it interval graph}. If these intervals are constructed such that no interval properly contains another then such graph is called a {\it proper interval graph}. The families of interval and proper interval graphs are widely studied and used in different fields. In this chapter we present an algorithm which produces an optimal solution of the MinLA on proper interval graphs.

\par Let us construct graph $G=(V,E)$ in a following way (algorithm $A$):
\begin{itemize}
  \item set $n$ as number of vertices in a graph
  \item drop $n$ vertices on an axis with integer coordinates from $1$ to $n$
  \item take a subset of successive vertices and make a clique from them
  \item return to the previous step $t$ times
\end{itemize}
\par
\begin{figure}[h]
  \vbox{\center\epsfig {figure=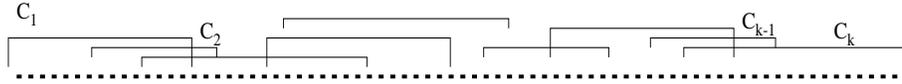,width=1cm,height=12cm,angle=270}}
  \caption{Schematic example of the chain of cliques}
  \label{schem}
\end{figure}

 As a result of this construction we obtain a graph with the representation like on Figure \ref{schem}. If we have a situation with nested cliques, we can ignore the clique that is placed inside of some other clique. We solve the problem for a family of graphs obtained by applying the algorithm $A$ and then show that there is an algorithm which produces such representation for proper interval graphs.
\par In the following claims we will work with a graph $G=(V,E)$ that is a chain of $k$ cliques $C_1...C_k$ constructed using algorithm $A$. In all following orders we index the vertices from $1$ to $n$, where $|V|=n$.

\par The orders of the vertices that preserve the order of cliques $C_1, C_2,...,C_k$ and full rotation $C_k,C_{k-1},...,C_1$ will be called 'natural orders' (or $N-order$) of $G$.
\par Denote by $d_v$ the degree of vertex $v$.

\section{The Algorithm}
Let us formulate and prove two claims that will serve as the basis of induction for Claim \ref{mainclaim}.

Claims \ref{b1} and \ref{b2} refer to a graph $G=(V,E)$ that is a union of two cliques $C_1$ and $C_2$ whose intersection is not empty and no clique contains the other (all other cases of union of two cliques are trivial).

\begin{claim}\label{b1}
  Given a graph $G=(V,E)$. If $G$ is a union of two cliques $C_1$ and $C_2$ then in every optimal linear order of $G$ the last vertex will not be from the intersection of $C_1$ and $C_2$.
\end{claim}
\begin{proof}
  Take any optimal order $\varphi$. Assume that the claim is false and the last vertex $u$ in optimal arrangement $\varphi$ comes from the intersection. Let us take vertex $v$ that is not in the intersection and $\varphi (v) = max_{x\not\in (C_1\cap C_2)}\varphi (x)$. Suppose that $\varphi (v)=n-k$ and w.l.o.g. $v\in C_1\setminus (C_1\cap C_2)$. Exchange $u$ and $v$. Since
\[
N(v)\subset N(u),
\]
by exchanging $u$ and $v$, the connections to $N(v)$ do not increase the cost of $\varphi$ after the flip. However $u$ is connected to the vertices from  $C_2\setminus (C_1\cap C_2)$ too. By moving $u$ to the position $n-k$ we reduce the cost of these edges, and does not change the cost of edges from $u$ to the vertices at $[n-k+1,...,n-1]$ positions (because they are from $C_1\cap C_2$). So $u\rightleftarrows v$ flip must reduce the cost of the optimal arrangement and this is a contradiction to the assumption.
\end{proof}

\begin{claim}\label{b2}
  Given a graph $G=(V,E)$. If $G$ is an union of two cliques $C_1$ and $C_2$ then every optimal linear order of $G$ has the following structure
  \[
  <\{v~|~v\in C_1\setminus (C_1\cap C_2)\},\{u~|~u\in (C_1\cap C_2)\},\{w~|~w\in C_2\setminus (C_1\cap C_2)\}>
  \]
  or
  \[
  <\{v~|~v\in C_2\setminus (C_1\cap C_2)\},\{u~|~u\in (C_1\cap C_2)\},\{w~|~w\in C_1\setminus (C_1\cap C_2)\}> 
  \]
  Call the order of this type $N_{2C}-order$.
\end{claim}

\begin{proof}
  Let us prove the claim by induction on a number of vertices in $G$.
  \begin{itemize}
    \item \par The basis of induction is when $|V|=1$ and the claim in this case is true.
    \item \par Suppose that if $|V|=n-1$ the claim is true.
    \item \par Let $|V|=n$. Take some optimal order $\varphi$ of $G$. Suppose, that w.l.o.g. the last vertex in $\varphi$ is $v\in C_1\setminus (C_1\cap C_2)$ (as was proved in Claim \ref{b1}). Remove $v$ from $G$ with all adjacent edges. We obtain a new graph $G^{'}$ with $n-1$ vertices. Look at its arrangement $\psi$ such that $\psi(u)=\varphi(u)$. There are two possible cases : $\psi$ is optimal and $\psi$ is not optimal.
      \par If $\psi$ is optimal then, by induction hypothesis, it has  $N_{2C}-order$ type and, after returning $v$ to $G$, we save an $N_{2C}-order$ type. Since we started with the optimal order then $\varphi$ has $N_{2C}-order$ type.
      \par Assume that $\psi$ is not optimal. Then there exists some optimal order $\rho$ for $G^{'}$ and $C_{G^{'},\psi} > C_{G^{'}, \rho}$. Thus
      \[
      C_{G, \varphi}\geq C_{ G^{'}, \psi}+\alpha > C_{G^{'}, \rho} + \frac{d_v(d_v+1)}{2}  
      \]
      Recall that $\varphi$ is optimal for $G$. But $C_{G^{'}, \rho}$ produces a cost of linear order which is $N_{2C}-order$ on $G^{'}$ and, by adding vertex $v$, we save $N_{2C}-order$ increasing $C_{G^{'}, \rho}$ by $\frac{d_v(d_v+1)}{2}$. So we obtain a strict inequality for given optimal order $\varphi$ and this is a contradiction. Thus $\psi$ cannot be not optimal.
    \end{itemize}
\end{proof}

\par Let us switch to the main claim.

\begin{claim}\label{mainclaim}
  Given a graph $G=(V,E)$ constructed using algorithm $A$ then every minimum linear order of $G$ is an N-order.
\end{claim}

\begin{proof}
\par Suppose that after removing all nested cliques, the graph will remain with $k$ different cliques. We can assume that $G$ is connected, otherwise the problem can be divided into the similar subproblems per connected component. Let us prove it by induction on $k$.
\begin{itemize}
  \item When $k=0$ or $1$ the claim is trivial and when $k=2$ we prove it in claims \ref{b1} and \ref{b2}.
  \item Suppose that the claim is true for $k-1$ and let us prove it for $k$ cliques chain. Here we use the second induction on the number of vertices in $T=C_k\setminus \cup_{i=1}^{k-1}C_i$ (call it $l$).
    \begin{enumerate}
    \item When $l=0$ there are $k-1$ cliques in $G$ and the claim is true by induction hypothesis.
    \item If $l=1$ there exist unique vertex $v$ in $T$. We will call vertex $u\in G$ $p$-vertex if it belongs to the intersection of $C_k$ with some other clique. Assume that we have some optimal order $\phi$. If $\phi$ is N-order then the claim is true. Suppose that $\phi$ is not an N-order. Then there exist two possible cases of $\phi$'s structure depicted in Figure \ref{twocases}. 

\begin{figure}[htb]
  \vbox{\center\epsfig {figure=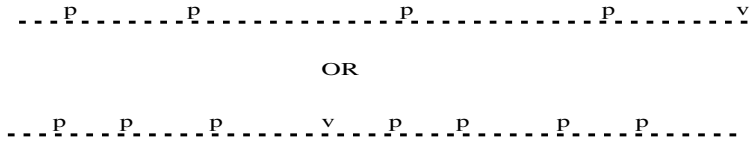,width=10cm,height=1.8cm}}
  \caption{Possible cases of structure of $\phi$}  
  \label{twocases}
\end{figure}

\par {\bf First case : }If $\phi(v)>\phi(p_i)$ (or less) for any $i$ then      
\[
C_{G,\phi}~=~C_{G\setminus v,\phi} + \alpha
\]
where $\alpha \geq \frac{d_v(d_v+1)}{2}$. From the other hand we have
\begin{equation}\label{norder}
C_{G,N-order}~=~C_{G\setminus v,N-order} + \frac{d_v(d_v+1)}{2}
\end{equation}
Then

\begin{align*}
C_{G\setminus v,\phi} + \alpha ~ \leq ~C_{G\setminus v,N-order} + \frac{d_v(d_v+1)}{2}\\
C_{G\setminus v,\phi} + \alpha  - \frac{d_v(d_v+1)}{2} ~ \leq ~C_{G\setminus v,N-order} \\
\end{align*}

So we obtain that $\phi$ is at least good at $G\setminus v$ as $N-order$. The situation when it is better is impossible by induction hypothesis and if it is not better then it is exactly $N-order$ and then on $G$ it will be also $N-order$.

\par {\bf Second case : }$v$ is placed between $p$-vertices. Suppose w.l.o.g. that there are $L$ $p$-vertices $p_i$ such that
\[
\forall ~ p_i ~  \phi(p_i)<\phi(v),~1\leq i \leq L
\]
and $R$ $p$-vertices $q_i$ such that
\[
\forall ~ q_i ~  \phi(q_i)>\phi(v),~1\leq i \leq R
\]
then
\[
C_{G,\phi}~ \geq ~ C_{G\setminus v,\phi} + \frac{R(R+1)}{2} +  \frac{L(L+1)}{2} + LR.
\]
From the other hand we have equation \ref{norder}. Combining them we obtain

\begin{multline}
C_{G\setminus v,N-order} + \frac{d_v(d_v+1)}{2}~= \\=~C_{G\setminus v,N-order} + \frac{(L+R)(L+R+1)}{2}~ \geq \\
\geq ~ C_{G\setminus v,\phi} + \frac{R(R+1)}{2} +  \frac{L(L+1)}{2} + LR
\end{multline}

and then
\begin{equation}\label{isstr}
C_{G\setminus v,N-order} \geq C_{G\setminus v,\phi}.
\end{equation}
If inequality \ref{isstr} is strict, this is a contradiction to the induction hypothesis and if there is an equality, $\phi$ should be an $N-order$ by induction. However, placing $v$ in the middle of the $N-order$ cannot allow the optimal order and this is a contradiction too.

\item Suppose the claim is true for $|T|=l-1$ and now we need to prove it for $l$. Let $v$ be the $l$-th vertex in $T$. Assume that $\phi$ is some optimal order. If $\phi$ is an $N-order$ then the claim is true. Suppose that $\phi$ is not an $N-order$. If $v$ is the last (or first, the same proof in this case) vertex in $\phi$ then
\[
C_{G,\phi}~=~C_{G\setminus v,\phi} + \alpha
\]
where $\alpha \geq \frac{d_v(d_v+1)}{2}$. From the other hand we have
\[
C_{G,N-order}~=~C_{G\setminus v,N-order} + \frac{d_v(d_v+1)}{2}
\]
Then

\begin{align*}
C_{G\setminus v,\phi} + \alpha ~ \leq ~C_{G\setminus v,N-order} + \frac{d_v(d_v+1)}{2}\\
C_{G\setminus v,\phi} + \alpha  - \frac{d_v(d_v+1)}{2} ~ \leq ~C_{G\setminus v,N-order}
\end{align*}

So we obtain that $\phi$ is at least good at $G\setminus v$ as $N-order$. If $\phi$ is better then it is impossible by the induction hypothesis. Otherwise, if it is not then $\phi$ is exactly $N-order$ and then on $G$ it will be also $N-order$.
\par Suppose that $v$ is not last (or first) and no other vertex from $T$ is (in this case we can flip them and obtain the same optimal order). Then there are two indexes $i$ and $j$ s.t. $1<i<j<n$, where $i$ is a first occurrence of some vertex from $T$ in $\phi$ and $j$ is a last. Let us flip $v$ with $j$-th vertex and call the new order $\phi'$. The cost of $\phi'$ is remained the same as it was in $\phi$ because the flipped vertices have the same set of neighbors. Like in a previous case, there are $L$ $p$-vertices to the left of $v$ and $R$ to the right. If $R=0$, we remove $v$ like in the case of $l=1$ and obtain a contradiction to the induction on $l$. Otherwise, if $R>0$ then 
\[
C_{G,\phi'}~ \geq ~ C_{G\setminus v,\phi'} + \frac{R(R+1)}{2} +  \frac{(l+L-1)(L+l)}{2} + (L+l-1)R.
\]
From the other hand we have equation \ref{norder}. Combining them we obtain

\begin{multline}
C_{G,N-order}~ =~ C_{G\setminus v,N-order} + \frac{d_v(d_v+1)}{2}~ =\\
=~C_{G\setminus v,N-order} + \frac{(l-1+L+R)(l+L+R)}{2}~ \geq~ \\
\geq~ C_{G,\phi'}~ \geq ~ C_{G\setminus v,\phi'} + \frac{R(R+1)}{2} +  \frac{(l+L-1)(L+l)}{2} + (L+l-1)R
\end{multline}

This case is similar to the case of $l=1$ but while removing $v$ we use the induction on $l$ and the claim is true.
\end{enumerate}
\end{itemize}
\end{proof}

Now we can define a process of calculation of the MinLA on proper interval graphs in polynomial time.

\begin{claim}
  The minimum linear order of proper interval graph is calculated in polynomial time.
\end{claim}
\begin{proof}
  Proper interval graphs can be recognized by a linear order: their vertices can be linearly ordered such that the vertices contained in the same clique are consecutive. The recognition works in linear time and produces such order \cite{lexbfs}. Graphs with such possible order may be created with algorithm $A$. Then the minimum linear arrangement of proper interval graphs is calculated in polynomial time.  
\end{proof}

\par Thus, the algorithm for calculation of the MinLA is following :
\begin{enumerate}
  \item Apply any polynomial algorithm for proper interval graph recognition which produces an $N-order$ of vertices,
  \item This order is a minimum linear arrangement.
\end{enumerate}

\section{4-Approximation for interval graphs}

\begin{figure}[h]
  \vbox{\center\epsfig {figure=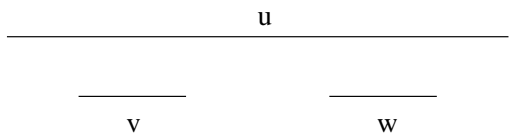}}
  \caption{Counterexample of interval graph.}
  \label{badgr}
\end{figure}

The complexity of the MinLA problem on general interval graphs is not known. The order of vertices by their start (or finish) time is not optimal already on the example at Figure \ref{badgr}.
In this chapter we introduce 4-approximation algorithm for the problem on interval graphs. Before we will start with the algorithm let us look at some preliminary facts :

\begin{theorem}
  {\bf (Gilmore-Hoffman [1964])} The following are equivalent :
  \begin{enumerate}
    \item $G$ is an interval graph.
    \item It is possible to order the maximal cliques in $G$ so that for every $v\in V$ the cliques that contain $v$ appear consecutively in this order.
  \end{enumerate}
\end{theorem}


The polynomial algorithms that construct such orders are introduced in many sources (for example in \cite{golumbic}) given an interval graph (possibly with no interval representation). Suppose we have such clique order $\phi$
\[
C_1, C_2, ... , C_k.
\]
Then for every vertex $v\in V$ it is possible to define a segment $[C_i,...,C_{i+l}]$ in $\phi$ that contains the maximal cliques of $v$. Let us call the corresponding $i$ and $i+l$ by $s_v$ and $f_v$ respectively (start and finish). Now we can order the vertices of the graph by their $s_v$ values (call the order by $\pi$ and ordered vertices by $v_1,...,v_n$).
\begin{theorem}
  Linear order $\pi$ is a 4-approximation for the Minimum Linear Arrangement problem on general interval graphs.
\end{theorem}
\begin{proof}
  Let us estimate the cost of arbitrary $\pi$-order. We call an incident to $u$ edge $e=(uv)$ in order $\pi$ -- ``right oriented edge'' if
  \[
  \pi(v)>\pi(u).
  \]
  The similar definition will be for ``left oriented edges''. Every vertex in $\pi$ has ``right oriented edges'' and/or ``left oriented edges'' as well. Every edge in $\pi$ is right oriented and left oriented for its different ends. We will pass vertex by vertex in order $\pi$ and estimate the total cost of right oriented edges for every vertex and this will give the estimation of the total cost of the order $\pi$. Denote by $R_j$ the set of right oriented edges of vertex $v_j$ in order $\pi$.
  \[
  C_{G,\pi} = \sum_{v_i \in V}\sum_{e\in R_i}cost_{\pi}(e).
  \]
Since the vertices are ordered by their $s_v$ values
\[
\sum_{e\in R_i}cost_{\pi}(e) \leq \frac{d_{v_i}(d_{v_i}+1)}{2}.
\]
Thus
\[
C_{G,\pi} \leq \sum_{v \in V}\frac{d_{v}(d_{v}+1)}{2}.
\]

From the other hand for any optimal order $\psi$ the following is true :
  \begin{equation}\label{opt}
  C_{\psi^*}~\geq ~ \frac{\sum\limits_{v} \biggl( 2 \frac{\frac{d_v}{2}(\frac{d_v}{2}+1)}{2} \biggr) }{2}~=~ \frac{\sum\limits_{v} \biggl( \frac{d_v}{2}(\frac{d_v}{2}+1) \biggr) }{2}.
  \end{equation}
  Combining previous results we obtain
  \begin{equation}
    A~=~\frac{\sum\limits_{v} \biggl( \frac{d_v}{2}(\frac{d_v}{2}+1) \biggr) }{2} ~\leq ~ C_{\psi^*} ~\leq ~ C_{\pi}~\leq~ \sum\limits_{v} \frac{d_v(d_v+1)}{2}~=~B.
  \end{equation}
  Clearly that for $\alpha=4$
  \[
  B\leq \alpha A
  \]
  and this proves that any $\pi$-order gives $4$-approximation for the Minimum Linear Arrangement problem on general interval graphs.
\end{proof}
\par Easy to see that given some interval representation of interval graph, the $\pi$-order is equivalent to the order by start (or finish) time of the intervals.

\end{document}